# Topological Magnetoresistance of Magnetic Skyrmionic Bubbles

Fei Li[1,2,3], Hao Nie[1,3], Yu Zhao[1,3], Zhihe Zhao[1,2], Juntao Huo[4], Hongxian Shen[1,3], Sida Jiang[5], Renjie Chen[2], Aru Yan[2], S-W Cheong[6], Weixing Xia[2]*, Lunyong Zhang[1,3]* and Jianfei Sun[1,3]

[1]. School of Materials Science and Engineering, Harbin Institute of Technology, Harbin 150001, China

[2.] CISRI & NIMTE Joint Innovation Center for Rare Earth Permanent Magnets, Ningbo Institute of Material Technology and Engineering, Chinese Academy of Science, Ningbo 315201, China

[3.] National Key Laboratory of Precision Hot Processing of Metals, Harbin Institute of Technology, Harbin 150001, China

[4.] Key Laboratory of Magnetic Materials and Devices, Ningbo Institute of Material Technology and Engineering, CAS, Ningbo 315201, China

[5.] Space Environment Simulation Research Infrastructure, Harbin Institute of Technology, Harbin 150001, China

[6.] Rutgers Center for Emergent Materials and Department of Physics and Astronomy, Rutgers University, 136 Frelinghuysen Rd, Piscataway, NJ, United States of America

Corresponding authors: W.X. Xia    xiawxing@nimte.ac.cn

L.Y. Zhang zhangly@hit.edu.cn; allen.zhang.ly@gmail.com

## Abstract

Magnetic skyrmions offer promising prospects for constructing future energy-efficient and high-density information technology, leading to extensive explorations of new skyrmionic materials recently. The topological Hall effect has been widely adopted as a distinctive marker of skyrmion emergence. Alternately, here we propose a novel signature of skyrmion state by quantitatively investigating the magnetoresistance (MR) induced by skyrmionic bubbles in CeMn$_2$Ge$_2$. An intriguing finding was revealed: the anomalous MR measured at different temperatures can be normalized into a single curve, regardless of sample thickness. This behavior can be accurately reproduced by the recent chiral spin textures MR model. Further analysis of the MR anomaly allowed us to quantitatively examine the effective magnetic fields of various scattering channels. Remarkably, the analyses, combined with the Lorentz transmission electronic microscopy results, indicate that the in-plane scattering channel with triplet exchange interactions predominantly governs the magnetotransport in the Bloch-type skyrmionic bubble state. Our results not only provide insights into the quantum correction on MR induced by skyrmionic bubble phase, but also present an electrical probing method for studying chiral spin texture formation, evolution and their topological properties, which opens up exciting possibilities for identifying new skyrmionic materials and advancing the methodology for studying chiral spin textures.

**Introduction**

Magnetic skyrmions are prospected as one of the basis for future information technologies like the potential racetrack memories and neuromorphic computing, due to their stable topological charges associated with chiral spin textures, small size and energy-efficiency.[1-2] One of the key steps to realize such device applications is reading out the information carried out by them, which requires to translate the chiral spin texture to readable electric signals.[3-6] The novel magneto-electric effects initialed by magnetic skyrmions thus attract much attentions.[3, 7-8] In conductive systems with chiral spin textures, the exchange interaction of itinerant electrons with non-colinear/coplanar spins will induce additional scattering potential of moving electrons, which generates unconventional magnetotransport behaviors like the so-called topological Hall effect (THE), an additional transverse resistance term beyond the normal and anomalous Hall contributions. THE is caused by a finite Berry curvature $\boldsymbol{\Omega} = \boldsymbol{m}\left(\frac{\partial \boldsymbol{m}}{\partial x} \times \frac{\partial \boldsymbol{m}}{\partial y}\right)$ associated with the chiral spin texture $\boldsymbol{m}$(x, y) in theory which applies on itinerant electrons as an effective magnetic field $\boldsymbol{B}_{eff} = \frac{1}{2}\boldsymbol{\Omega}$, as schematically shown by Fig.1.[9] It has been therefore deemed as an electric signature of chiral spin textures and was always used as an phenomenological probe of magnetic skyrmions recently.[6, 10-12]

On the other hand, it is natural to expect that the effective magnetic field emerged from the chiral spin texture Berry curvature would simultaneously modify the longitudinal resistance (see Fig.1), analogy to the origination of traditional Lorentz force inducing magnetoresistance (MR). A correction to MR thus should be there along with THE as proposed by the recently developed interference-induced quantum correction theory of MR induced by chiral spin textures[13], which is distinct from the traditional MR in non-chiral spin states. The MR correction is governed by the size and magnetization spatial distribution of chiral spin textures $\boldsymbol{m}$(x, y), consequently provides another electric signature of chiral spin textures, in particular, skyrmions. However, such nontrivial MR effect were not experimentally validated up to date. A few previous works reported MR anomaly in some skyrmion systems such as MnSi[14-16], FeGe[17] etc. and argued that the anomaly was relevant to skyrmion formation because they were observed in a corresponding magnetic field region. The anomaly was often arbitrarily ascribed to spin fluctuation scattering during the skyrmion phase evolution.[18] This scenario is distinct from the mechanism of skyrmions causing MR since spin fluctuations are not relevant to the topology of chiral spin texture.

The present work reports significant skyrmionic bubble inducing MR in CeMn$_2$Ge$_2$, which can be simulated using the interference-induced quantum correction theory for chiral spin textures[13]. CeMn$_2$Ge$_2$ is a member of the RMn$_2$Ge$_2$ (R = rear earth) compounds family with complex non-collinear spin arrangement[19-23]. Room temperature skyrmionic bubbles state was probed recently in the polycrystal sample of this compound[24-25] as well as its counterparts LaMn$_2$Ge$_2$[26] and NdMn$_2$Ge$_2$[27-28], by large THE and/or Lorentz transmission electronic microscopy (LTEM). Here we investigated the magnetoresistance behaviors of CeMn$_2$Ge$_2$ single crystal samples with varied thicknesses in details. Large MR anomaly was observed on the negative MR background in the magnetic field region corresponding to skyrmionic bubble phase formation revealed by THE and LTEM. It was quantitatively simulated by the chiral spin textures MR model, which indicates that the Bloch-type skyrmionic bubbles support a significantly stronger in-plane scattering channel with triplet exchange interaction to the itinerant carriers. Magnetoresistance anomaly following the chiral spin textures MR model is thus acting as a quantitative signature of the skyrmionics bubble formation and evolution, in analogy to THE, and could be termed as topological magnetoresistance.

Experimental details

Single crystals of CeMn$_2$Ge$_2$ were grown using an indium flux method. Ce lumps (purity 99.9%), Mn powders (purity 99.95%), Ge powders (purity 99.99%) and In powders (purity 99.99%) with a molar ratio of Ce: Mn: Ge: In = 1:2:2:30 were sealed in a quartz ampoule under high vacuum. The sealed quartz ampoule was firstly heated up to 1373 K at a rate of 100 K/h and held for 24 h, and following cooled down slowly to 973 K at a rate of 3 K/h. Finally, the ampoule was taken out from the furnace and centrifugally decanted to separate the single crystals from residue flux.

X-ray diffractometer with Cu-$K_\alpha$ radiation ($\lambda$ = 0.15406 nm) was used to collect the powder XRD pattern of the samples at room temperature for determining their phase purity and crystal structure. Rietveld refinement of the XRD spectra was carried out with the FullProf software package. The specimens were cut and polished into rectangle shape for further magnetization and electrical transport measurements. The polish treatment removed the small amount residual indium flux on the single crystal surface. Two samples with thickness of 0.65 mm and 0.11 mm were prepared for the electrical transport measurements.

Magnetization and electronic transport properties were measured by a Quantum Design Physical Property Measurement System (PPMS-14 T). The six-probes method was used to measure the longitudinal ($\rho_{xx}$) and Hall resistivities ($\rho_{xy}$) on the electronic transport option (ETO) module. The current was applied along the length direction ($I // a$) and the magnetic field along the $c$-axis ($H//c$) during the electronic transport measurements. Symmetrizing or antisymmetrizing the measurements with positive and negative field were adopted to remove the small amounts of mixing between $\rho_{xx}$ and $\rho_{xy}$ due to slight misalignments of electrodes. The magnetic domain structure was detected by using a Talos F200x machine in the LTEM mode, where the fields were applied by objective lens. The thin plates for LTEM observation were prepared by traditional mechanical polishing and argon ion milling.

## Results and Discussion

CeMn$_2$Ge$_2$ crystallizes into a ThCr$_2$Si$_2$-type tetragonal structure of centrosymmetric space group *I4/mmm*, as shown in Fig.2a. The magnetic Mn layers are separated by the Ce and Ge layers[25, 29], which makes the single crystal samples exhibiting layered compounds morphology as thin plate, see the inset photograph in Fig.2b. The samples are naturally cleavable with the large square surface as the *ab*-plane. Figure 2b displays the XRD pattern of crushed powders of the single-crystalline specimens. Rietveld refinement analyses of the pattern indicated high purity and crystallinity single-phase CeMn$_2$Ge$_2$ with right crystal structure was obtained. The refined lattice parameters are $a = b = 4.144(1)$ Å, $c = 10.929(8)$ Å, in coincidence with the reported values[25, 30].

Figure 2c shows the magnetization as function of temperature (MT) measured under zero-field-cooled (ZFC) and field-cooled conditions (FC) with an external field ($\mu_0H$=500 Oe) paralleling to the *c*-axis and *a*-axis respectively. Strong magnetic anisotropy is observed with the *c*-axis as the easy magnetized direction, as indicated as well by the field dependent magnetization (MH, see Fig.2d) where the saturated field (~0.18 T) for *H//c* is about 5 times smaller than that (~1.0 T) for *H//a* at 300 K. The rest of isothermal magnetization curves for both field configurations at various temperatures (Figure S1, Supporting Information) confirm the large magnetic anisotropic with the easy *c*-axis. This uniaxial easy axis anisotropy, competing with the magnetic dipole interaction, produces skyrmion bubbles in the present investigated CeMn$_2$Ge$_2$ compound with a centrosymmetric crystal structure. Moreover, it could be concluded that the magnetic moments at ground state are inclined to align with the *c*-axis, corresponding to the reported low temperature Mn-intralayer conical ordered spin state with a spin propagation vector along the *c*-axis[31]. This conical spin phase would evolve into an antiferromagnetic state forming in the Mn-interlayers at around $T_{\text{inter}} = 320$ K as temperature increasing through a mediated Mn-interlayer canting ferromagnetic state[21, 30, 32]. The skyrmions bubble phase is observed in the transition region[24].

Our samples consistently hold a magnetic phase transition region starting at 322 K as temperature lowering, as demonstrated by the magnified figure of the transition region (see Fig.S2a in the supporting information). The magnetization sharply increased at 322 K determined by the first inflection points of the differential of magnetization, d$M$/d$T$. (Fig. S2b in the supporting information).

We furthermore observed a magnetization kink very close to the transition point 322 K ( Fig.S2a in the supporting information) which was not reported in literatures before. It was occurred at ~317 K, as determined by the second inflection points on the d$M$/d$T$ curves (Fig. S2b in the supporting information) and could be ascribed to a spin reorientation process. Such spin reorientation behaviors were often reported in materials hosting topological magnetic structures[33-36]. The magnetization of $H$//a was then steeply decreased from $T_{intra}$ = 316 K (Fig.2c and Fig.S2a). This intensive spin flipping behavior was deemed as the transition process from a Mn-interlayer canted ferromagnetic state to the Mn-intralayer conical spin state, where the canted ferromagnetic state occurred in the narrow region from 316-322 K[31, 37].

The longitudinal resistivity under zero magnetic field have been measured as a function of temperature $T$, see Fig.2e. A typical metallic behavior has been detected over the entire temperature range. The residual resistivity ratio (RRR) calculated by $\rho_{xx}$(300 K)/$\rho_{xx}$(2 K) is ~24.04, indicating the high quality of the as-grown single-crystalline specimens. An inflection point noted by the triangle symbol is observed at 322 K corresponding to the magnetic phase transition point $T_{inter}$ probed in MT, indicating strong magnetic scattering of electrons in this compound because the crystallographic structure is stable in the temperature range.

We furtherly measured the samples' Hall resistivity $\rho_{xy}$ as shown in Figs.S3a and S4a (Supporting Information) respectively for the 0.11 mm thick specimen and the 0.65 mm thick specimen. The measurement geometry was schematically depicted as Fig.S3c. The $\rho_{xy}$ curves exhibit a strong resemblance to the MH ones, suggesting a domination of anomalous Hall effect (AHE) due to the field driven magnetization. Here the following total Hall resistivity equation of a magnet was used to quantitatively analyze the $\rho_{xy}$ data[28]:

$$\rho_{xy}(\mu_0 H) = \rho_{xy}^N + \rho_{xy}^A + \rho_{xy}^T = R_0 \mu_0 H + R_s M(\mu_0 H) + \rho_{xy}^T \qquad (1)$$

where $\rho_{xy}^N, \rho_{xy}^A$, $R_0$ and $R_s$ are the ordinary and anomalous Hall resistivity, ordinary and anomalous Hall coefficients respectively, $\rho_{xy}^T$ is the THE resistivity. Then, it was indicated that THE components $\rho_{xy}^T$ were necessary to account for the probed $\rho_{xy}$ data, for example see Fig.S3b (Supporting Information). The $\rho_{xy}^T$ at various temperatures were summarily shown in Fig.2f and Fig.S4b (Supporting Information) for the 0.11 mm thick sample and 0.65 mm thick sample, respectively. Obvious negative $\rho_{xy}^T$ exist from room temperature to at least 100 K below the saturated fields, which suggests the formation of chiral spin structure in our samples. Figure 4e and Fig. S4d constructed the field-temperature magnetic phase diagram for the 0.11 mm and 0.65 mm thick sample respectively based on the field dependent THE resistivity $\rho_{xy}^T$ summarized in Fig.2f and Fig. S4b. Obviously, the

THE indicated a chiral spin state space spanning a wide temperature range about 175 K including room temperature.

To confirm the structure of the chiral spin state producing the THE, LTEM was used to directly image magnetic domain structures of our samples in real space. Fig.3a-3f and Movie S1(the starting 11 seconds, supporting information) display the evolution of magnetic domains recorded at 300 K under vertically applied magnetic fields along the *c*-axis. Labyrinthine domain structures could be seen under zero field, which are induced by the competition between magnetic anisotropy and dipolar-dipolar interaction and usually observed in centrosymmetric skyrmionic magnets[33, 38-40]. An alternate bright and dark contrast of the labyrinthine domain edge reveals the domain walls as Bloch type, indicating that the magnetization of the domains preferred to be aligned with the *c*-axis in CeMn$_2$Ge$_2$ flake sample. It should be emphasized that the stripe domains (pointed by the yellow dashed arrows) shrunk gradually upon field increase and evolved into an isolated skyrmionic bubble at ~220 Oe (Fig.3c). Figure 3g gives out the magnified picture of the bubble. Then, it kept shrinking inward and eventually disappeared at ~1100 Oe (Fig.3d-3f). The in-plane magnetization distribution of the magnetic bubble was resolved via the transport of intensity equation (TIE) analysis, as shown in Fig.3h, where the region with diffraction perturbation was masked by a white block and white arrows pointed out the directions of the in-plane magnetization projection, while the dark color indicates the almost out-of-plane magnetization. It was unveiled that the in-plane projection of magnetization vectors formed a counterclockwise type configuration, and the spin arrows at the center or background are negligible to show the magnetization there is out-of-plane. Such a configuration suggests that the observed bubble is topologically equivalent to a Bloch type skyrmion with topological number $Q = -1$[9]. Thus, the observed $\rho_{xy}^T$ dependent on applied magnetic field was induced by the skyrmionic bubbles formation and evolution in our samples, which ensured the nontrivial origination of the MR anomaly demonstrated following.

The MR was measured together with THE, with electrical current along the *a*-axis and magnetic field along the *c*-axis. Figure 4a and 4b show the MR defined as $[(\rho(H) - \rho(0))/\rho(0)] \times 100\%$ of the 0.11 mm and 0.65 mm thick specimens, respectively, at various temperatures. The specimens exhibited similar negative MR patterns, slow decrease at small fields ($\mu_0 H < \mu_0 H_c^{MR}$) and rapid decrease at larger fields, thus creating shoulders on the MR curves. The MR decreasing at high field ($\mu_0 H > \mu_0 H_c^{MR}$) followed a linear type trace, this behavior reminds the forced ferromagnetic (FM) state often observed in conventional FM materials. The transition between the two stages is abrupt, as seen more clearly from the differential curves of the MR where the slop of MR is dropped almost discontinuously at the critical field $\mu_0 H_c^{MR}$, especially at temperatures approaching the transition points $T_{intra}$ from canted ferromagnetic state to conical spin state, see Fig.4c and Fig.S4c (Supporting information) for the 0.11 mm and 0.65 mm thick samples respectively. Such steep change of MR slope suggests different phase states were founded in the two stages. This was elaborated by the loop scanning of MR and Hall resistance for the 0.11 mm sample at 300 K, shown in Fig.4d, with applied magnetic field from zero to 0.5 T and then inversely to zero tesla. Distinct but narrow hysteresis was observed for both the MR and $\rho_{xy}$ at the transition region. Same results were revealed in the thicker specimen as well (Fig.S5, Supporting Information). Figure 4d simultaneously demonstrated that the transition field of MR was in coincidence to the field at which the $\rho_{xy}$ reached saturation. We furtherly put the critical fields $\mu_0H_c^{MR}$ at different temperatures of the 0.11 mm and 0.65 mm thick specimens, respectively in the magnetic phase diagrams shown in Fig.4e and Fig.S4d. It was shown that the $\mu_0H_c^{MR}$ fields excellently overlap the $\mu_0H_c^{Hall}$ fields. The domain structure evolution with sweeping the magnetic field from high to low obtained by the LTEM video (the ending 27 seconds in Movie S1, Supporting Information) further demonstrated that the sharp drop of MR and $\rho_{xy}$ as $\mu_0H$ decreases is caused by the abrupt nucleation of labyrinthine type domains (at the 33-34s in Movie S1). Thus, the emergence of MR anomaly is perfectly accompanying with the THE, and associated with the formation of the skyrmionic bubbles.

Interestingly, the anomalous MR at various temperatures for the specimens of both thicknesses were converged into an unified curve, as shown in Fig.4f, by defining a normalized MR as $(\rho - \rho_c)/(\rho_0 - \rho_c)$ with respect to a normalized magnetic field $\mu_0H / \mu_0H_c^{MR}$, where $\rho_c$ notes the magneto-resistivity at the critical field $\mu_0H_c^{MR}$. However, the normalized MR at magnetic fields beyond $\mu_0H_c^{MR}$ diverged. Such a behavior again suggests different phase states in the two magnetic field regions and they support varied itinerant electrons scattering behaviors. Quantitatively, within the same definition of normalized MR and magnetic field, the so-called one-dimensional chiral soliton lattice was reported to host a normalized MR fallen into a unified curve under small field but divergence under large field, where the MR anomaly is induced by the conductive electrons scattering on the chiral spins of magnetic soliton kinks. Therefore, the anomalous MR we observed seemly is caused by a physical mechanism similar to that giving rise to the chiral soliton lattice MR, is an intrinsic property of the skyrmionics bubbles.

The unified MR curve of chiral soliton lattice can be described by the effective one-dimensional chiral sine-Gordon model of soliton density $L(0)/L(\mu_0H)=(\pi^2/4)K(\kappa)E(\kappa)$[41-42]. $K(\kappa)$ and $E(\kappa)$ denote the first and second kinds of complicate elliptic integrals, $\kappa$ is the elliptic modulus given by $\kappa/E(\kappa) = \sqrt{\mu_0H/\mu_0H_c}$ and $\kappa \in [0,1]$. $\mu_0H_c$ corresponds to $\kappa=1$. The model produces a reference curve as the black broken curve in Fig.4f. It was noted that the converged normalized MR of our samples does not follow the soliton density model in magnitude, indicating the skyrmionic bubble state is not equivalent to the soliton lattice state and it is possible to develop a novel unified model of skyrmion density with field evolution. Such difference in the normalized MR magnitude might be a result of the different dimensionality between the soliton lattice state and the skyrmion state. The front one is one-dimensional, but the skyrmion state is two-dimensional at least. Following this, the unified model of skyrmion density would be a two-dimensional version of the chiral sine-Gordon model.

Alternately, we proved that the anomalous MR can be quantitatively simulated by using the recent proposed interference-induced quantum correction theory of chiral spin textures MR[13]. The theory suggests that the conductivity correction for a two-dimensional chiral spin texture system under a weak perpendicular magnetic field $B=\mu_0H$ could be written as three possible models dependent on the specific scattering state of spin electrons, see supporting information for details. Here it was demonstrated that the anomalous MR in the low field regime could be well fitted by the following model Eq.2 (also the Eq.S1 in Supporting Information) in the case of coherent subbands scattering, for example realized in isolate skyrmionic bubbles. Such a coherent subbands scattering case is realized when the difference of Fermi wave vectors of the spin-up subband and spin down subband electrons $|k_{F1}-k_{F2}|$ are much smaller than the mean free path $\ell_{1,2}$ within each spin subband [13] (Supporting Information), this is satisfied in CeMn$_2$Ge$_2$ as an antiferromatic state (Fig.1c) founded in it which gives rise to very small splitting $\Delta$ between the spin-up and spin-down subbands, and CeMn$_2$Ge$_2$ is a good conductor (Fig.1e) with large $\ell_{1,2}$.

$$\Delta\sigma(B) = \frac{e^2}{2\pi h}\left[2f_2\left(\frac{B}{B_{t1}}\right) + f_2\left(\frac{B}{B_{t0}}\right) - f_2\left(\frac{B}{B_s}\right)\right] \qquad (2)$$

where $f_2(x) = \psi(1/2 + 1/x) + \ln x$, with the digamma function $\psi$, the magnetic fields $B_i$ ($i$=s, t0, t1) characterize the effective magnetic field applying on itinerant electrons. The s, t0 and t1 denote the singlet state, the triplet states with angular momentum projection as 0 and 1, respectively. Figures 5a and 5b gives out the fitted results for the 0.11 mm and 0.65 mm thick sample respectively, where the $\sigma(B)$ was obtained according to the relationship $\sigma(B) = \rho_{xx}(B)/[\rho_{xx}^2(B) + \rho_{xy}^2(B)]$. Thus, a consistent conclusion can be drawn from the analyses results of THE, LTEM and theoretical model simulations that the MR anomaly observed in our samples is a signature of the skyrmionic bubble state formation.

According to the MR fitting by Eq.2, we now can quantitatively analyze the effective magnetic fields $B_i$ ($i$=s, t0, t1) contributed by varied scattering channels which is proportional to the Berry curvature generated by chiral spin structure and has never been experimentally studied in detail before. The $B_i$ ($i$=s, t0, t1) were extracted out and plotted in Fig.5c. The triplet states with angular momentum projection as 1 expresses an effective magnetic field much larger than the other two, $B_{t1} > B_{s,t0}$, and the other two are almost same in magnitude, regardless of the sample thickness. That is to say, the electronic scattering is dominated by the triplet states channel with angular momentum projection 1, i.e. $|\uparrow\uparrow\rangle$ or $|\downarrow\downarrow\rangle$. In a word, the scattering does not cause spin flipping. This validates the spin conservation feature during the skyrmion scattering of electrons owing to Berry phase[43-44]. Then, because the spin relaxation caused by the $t1$ triplet states exchange interaction is suppressed in the case of a spin texture with only out-of-plane spin momentum[13], the dominant $t1$ scattering channel here implies that the electronic scattering is mainly induced by the interaction between the electronic spin and the in-plane spin momentum components of the skyrmionic bubbles. On the other hand, $B_i(i=t1,t0,s)$ almost keep increasing as temperature lower down and reach a maximum at a temperature point, this is corresponding to the evolution trend of $\rho_{xy}^T$ with temperature decrease, see Fig.2f and S4b, recalling the relationship of THE with the Berry curvature. Bearing in mind the Berry curvature is dependent on the geometrical configuration of skyrmions, the temperature dependence of the effective magnetic fields also gives a hint on how the skyrmion structure evolves in thermal field. Unfortunately, a clear function between the effective magnetic fields and skyrmion geometry such as its size is still unavailable so far.

## Conclusions

In summary, we observed skyrmionic bubble magnetoresistance and demonstrated its topological nature, directly relevant to the formation of chiral spin textures, which is analogous with the topological Hall effect. High-quality single-crystalline $CeMn_2Ge_2$ specimens with varied thicknesses were synthesized and a significant magnetoresistance was observed for all samples in a small magnetic field regime where the topological Hall effect and skyrmionics bubble phase were correspondingly observed, indicating the direct correlation between skyrmionic bubbles and the anomalous magnetoresistance. Furtherly, it was shown that the magnetoresistance could be normalized into one single curve regardless of the sample thickness by defining the normalized magnetoresistance $(\rho - \rho_c)/(\rho_0 - \rho_c)$ with respect to a normalized magnetic field $\mu_0 H / \mu_0 H_c^{MR}$, suggesting a hidden universality depicting the magnetoresistance induced by skyrmionic bubbles. We further proved that the anomalous magnetoresistance could be well reproduced by the interference-induced quantum correction theory for chiral spin textures[13]. Thus, the effective magnetic fields associated to the scattering channels of spin singlet and triplet states can be extracted out and investigated. The results proved the dominant spin-flipping-free scattering mode of skyrmionic bubbles, and the temperature dependent evolution of the effective magnetic fields as well as the skyrmionic structure. The present work therefore unveils the magnetoresistance anomaly as a probe of chiral spin textures and their topological properties, and sheds new light on the searching of new materials and phenomena of chiral spin textures such as skyrmions.

## Acknowledgment

We thank Professor W.D. Fei of Harbin Institute of Technology and Professor D. Wu of Nanjing University for enlightening discussions. The project is supported by the Funds of National Key Laboratory for Precision Hot Forming of Metals (JZKJW20220018-05), the "Head Goose" project of Heilongjiang Province (No. XNAUEA5640208620, XNAUEA5640200920-06), the Fundamental Research Funds for the Central Universities (DCQQ2970101922). W.X. Xia acknowledges the financial assistance from the Ningbo Key Scientific and Technological Project (Grant No. 2021000215)

## Author Contributions

L.Y. Zhang conceived the idea and designed the experiments. L.Y. Zhang, W.X. Xia and J.F Sun supervised the project. F. Li performed crystal growth, characterization, magnetization measurements, electrical transport measurements with the help from H. Nie, Y. Zhao, S.D. Jiang and J.T. Huo. F. Li performed the LTEM experiments and analyzed the images under the supervision of W.X. Xia and help from Z.H. Zhao, R.J. Chen and A. Yan. F. Li, L.Y. Zhang, H.X. Shen and J. T. Huo analyzed the data of crystal, magnetization and electrical transport measurements, F. Li, W.X. Xia, R.J. Chen and A. Yan analyzed the LTEM data. F. Li and L.Y. Zhang made the figures and wrote the manuscript with the critical reading and suggestions from S.W. Cheong. All authors contributed to the discussion of the results and improvement of the manuscript.

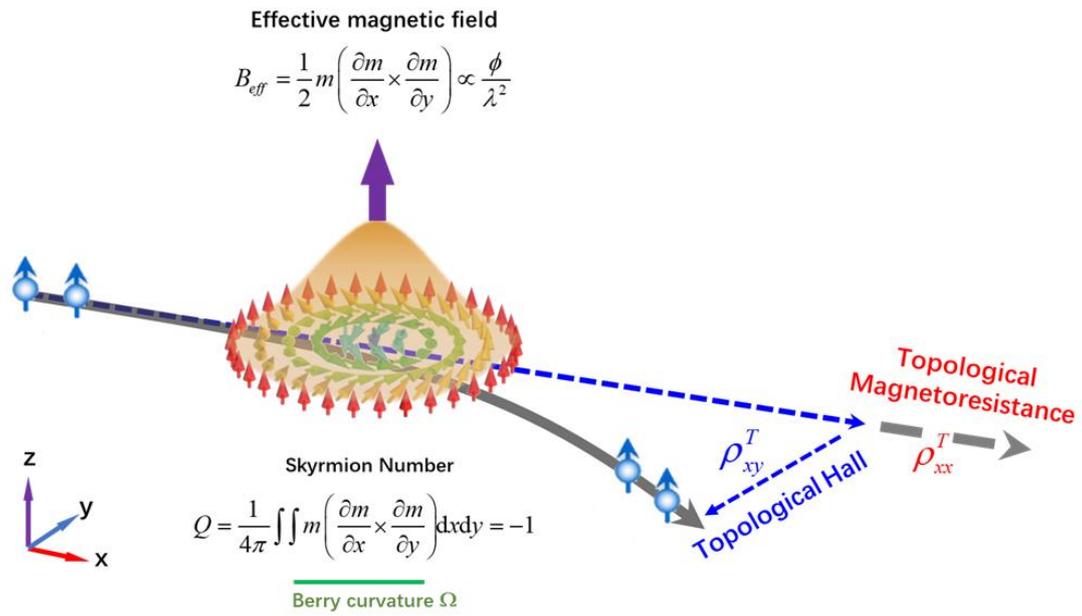

**Figure1** Schematic of electron motion as they travel through a skyrmion. Topological transport phenomena caused by a finite Berry curvature $\Omega = m\left(\frac{\partial m}{\partial x} \times \frac{\partial m}{\partial y}\right)$ associated with the chiral spin texture $m$(x, y) including the topological Hall effect and magnetoresistance correction effect.

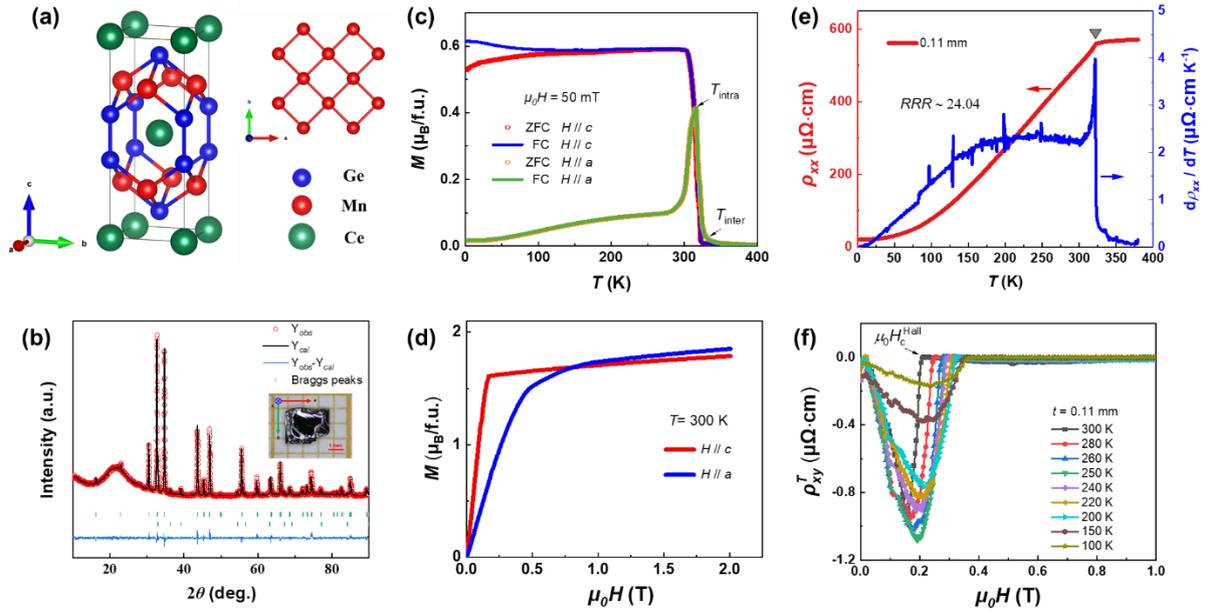

**Figure 2** (a) Sketch of CeMn$_2$Ge$_2$ crystal structure. (b) Rietveld refined XRD patterns of the crushed powder from single crystal samples, the inset shows a photograph of the as-grown single crystal. (c) Temperature dependence of magnetization in the ZFC and FC modes under 500 Oe fields for $H \parallel c$ and $H \parallel a$ respectively. The point of $T_{inter}$ and $T_{inter}$ are marked as black arrows. (d) The comparison of magnetization at 300 K between the case of $H \parallel c$ and $H \parallel a$. (e) The longitudinal resistivity and the first-order derivation of it as a function of temperature at zero field for 0.11mm thickness crystals and (f) Magnetic field dependence of topological Hall resistivity ($\rho_{xy}^T$) at different temperatures extracted from $\rho_{xy}(\mu_0 H)$ curves in Figure S3a (Supporting Information).

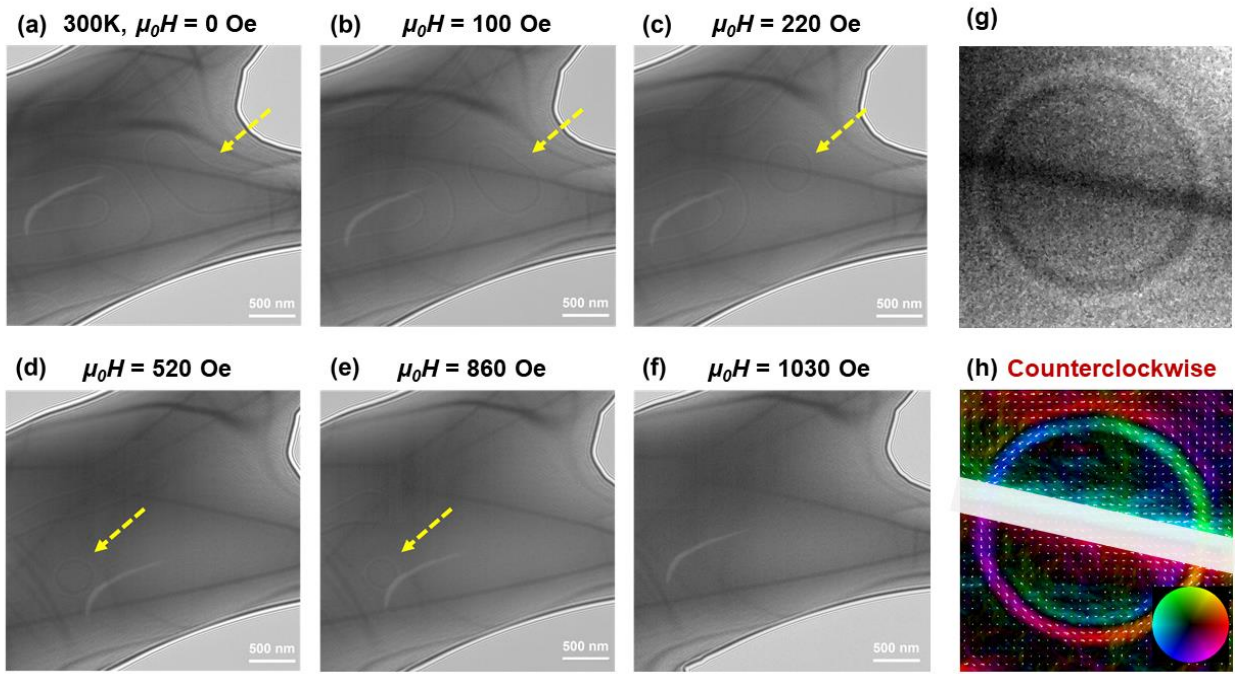

**Figure 3** (a)-(f) Evolution of the spin textures under vertically applied magnetic fields ($H /\!/ c$) at 300 K. All images are taken with a defocus value of −1.1 mm. The yellow dashed arrows show the magnetic domain evolved into a skyrmionic bubble with increasing field. (g) Enlarged LTEM image of the magnetic bubble in (c). (h) is the corresponding spin textures for the bubble extracted using TIE method (The region with diffraction perturbation was masked by a white block).

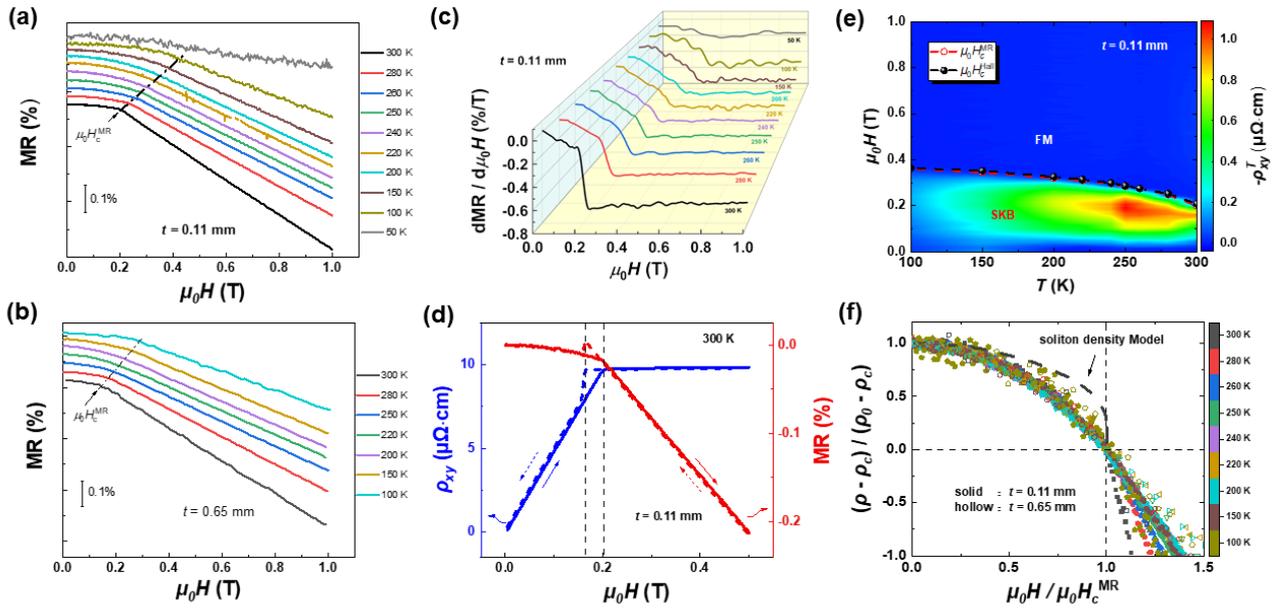

**Figure 4** (a) and (b) are the field dependence of longitudinal resistivities for the 0.11 mm and 0.65 mm thick sample respectively at different temperatures, black arrow indicates the critical field $\mu_0H_c^{MR}$ and the dotted line is a guide for eyes. (c) Derivative of the longitudinal resistivities data in (a) for the 0.11 mm thick sample. (d) The field dependence of magnetoresistance and Hall resistivity for the 0.11 mm thick sample taken with the same experimental procedure at 300 K for $\mu_0H /\!/ c$. The solid lines represent the measurement process with the field sweeping from 0 T to 0.5 T whereas the dash lines mean the field sweeps from 0.5 T to 0 T. (e) The magnetic phase diagram of the 0.11mm thick sample. The black and red solid circles correspond to the critical fields for the maximum absolute values extracted from Hall and MR curves respectively. (f) Normalized MR curves in a temperature region for both different thickness crystals (full symbols for the 0.11 mm thick sample and hollow symbols for the 0.65 mm thick sample respectively) and the black broken curve is acquired by the soliton density model $L(0)/L(\mu_0H)=(\pi^2/4)K(\kappa)E(\kappa)$.

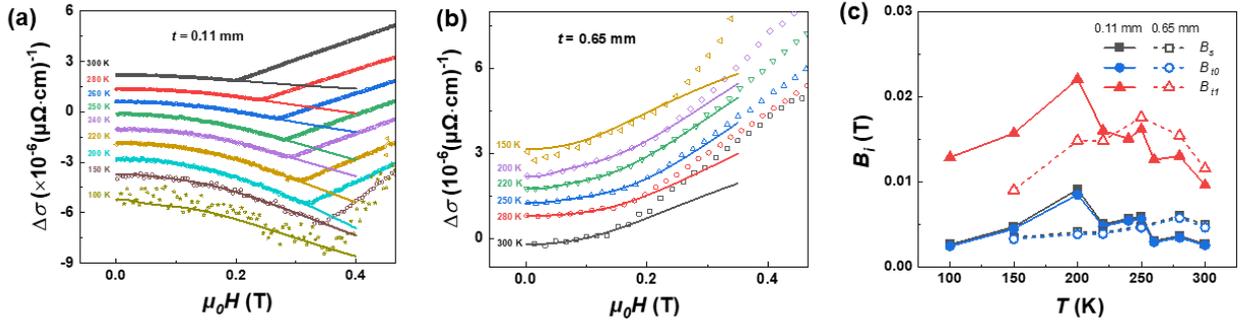

**Figure 5** (a) and (b) are magneto-conductance curves of the CeMn$_2$Ge$_2$ single crystal samples with 0.11 mm and 0.65 mm thicknesses at different temperatures respectively. Hollow points are experimental data and the solid lines are the corresponding fitting lines by the interference-induced quantum correction model Eq.2 of chiral spin textures MR. (c) The parameters $B_i(i=t1,t0,s)$ obtained by fitting for the samples of different thicknesses at measured temperatures (solid line for the 0.11 thick specimen and dot line for the 0.65 mm thick specimen).